\documentstyle[floats,aps,aipbook]{revtex}
\def\.{.\enskip}

\def\IntL{\ifmmode
              {\int{\cal L}dt}
           \else
              {\mbox{$\int{\cal L}dt$}}
           \fi}
\def\ee{\mbox{$e^+e^-$}}
\def\ppbar{\mbox{$p \bar p$}}
\def\to{\ifmmode{\rightarrow}\else{$\rightarrow$\enskip}\fi}

\def\micron{\ifmmode
                   {\mu{\rm m}}
               \else
                    \mbox{$\mu{\rm MeV}$}
               \fi}
\def\MeV{\ifmmode
                   {\rm \ MeV}
               \else
                    \mbox{${\rm MeV}$}
               \fi}
\def\GeV{\ifmmode
                   {\rm \ GeV}
               \else
                    \mbox{${\rm GeV}$}
               \fi}
\def\TeV{\ifmmode
                   {\rm \ TeV}
               \else
                    \mbox{${\rm TeV}$}
               \fi}
\def\invnb{\ifmmode
                   {{\rm \ nb}^{-1}}
               \else
                    \mbox{${\rm nb}^{-1}$}
               \fi}
\def\invpb{\ifmmode
                   {{\rm \ pb}^{-1}}
               \else
                    \mbox{${\rm pb}^{-1}$}
               \fi}
\def\D0{D$\emptyset$}
\def\pt{\ifmmode{p_T}\else{$p_T$\enskip}\fi}
\def\pl{\ifmmode{p_l}\else{$p_l$\enskip}\fi}
\def\plong{\ifmmode{P_{\parallel}}\else{$P_{\parallel}$\enskip}\fi}
\def\pperp{\ifmmode{P_{\perp}}\else{$P_{\perp}$\enskip}\fi}
\def\ptot{\ifmmode{P_{tot}}\else{$P_{tot}$\enskip}\fi}
\def\ptotal{\ifmmode{{\rm P}_{total}}\else{${\rm P}_{total}$\enskip}\fi}
\def\px{\ifmmode{{\rm P}_x}\else{${\rm P}_x$\enskip}\fi}
\def\py{\ifmmode{{\rm P}_y}\else{${\rm P}_y$\enskip}\fi}
\def\pz{\ifmmode{{\rm P}_z}\else{${\rm P}_z$\enskip}\fi}
\def\Et{\ifmmode{E_T}\else{$E_T$\enskip}\fi}
\def\Etot{\ifmmode{E_T^{tot}}\else{$E_T^{tot}$\enskip}\fi}
\def\Missing#1#2{{\mbox{$#1\kern-0.57em\raise0.19ex\hbox{/}_{#2}$}}\ }

\def\vMissing#1#2{\ifmmode
            \vec{#1}\kern-0.57em\raise.19ex\hbox{/}_{#2}
         \else
            {{\mbox{$\vec{#1}\kern-0.57em\raise.19ex\hbox{/}_{#2}$}}\ }
         \fi}

\def\phot{\ifmmode{\gamma}\else{$\gamma$\enskip}\fi}
\def\W{\ifmmode{W^{\pm}}\else{$W^{\pm}$\enskip}\fi}
\def\Z{\ifmmode{Z^\circ}\else{$Z^\circ$\enskip}\fi}
\def\Higgs{\ifmmode{}\else{$\H^\circ$\enskip}\fi}
\def\gluino{\ifmmode{\tilde{g}}\else{$\tilde{g}$\enskip}\fi}
\def\squark{\ifmmode{\tilde{q}}\else{$\tilde{q}$\enskip}\fi}
\def\slepton{\ifmmode{\tilde{l}}\else{$\tilde{l}$\enskip}\fi}
\def\sneutrino{\ifmmode{\tilde{\nu}}\else{$\tilde{\nu}$\enskip}\fi}
\def\stop{\ifmmode{\tilde{t}}\else{$\tilde{t}$\enskip}\fi}
\def\Zone{\ifmmode{\tilde{Z_1}}\else{$\tilde{Z_1}$}\fi}
\def\Ztwo{\ifmmode{\tilde{Z_2}}\else{$\tilde{Z_2}$}\fi}
\def\Zthree{\ifmmode{\tilde{Z_3}}\else{$\tilde{Z_3}$\enskip}\fi}
\def\Zfour{\ifmmode{\tilde{Z_4}}\else{$\tilde{Z_4}$\enskip}\fi}
\def\Wone{\ifmmode{\tilde{W_1}}\else{$\tilde{W_1}$}\fi}
\def\Wtwo{\ifmmode{\tilde{W_2}}\else{$\tilde{W_2}$\enskip}\fi}
\def\ee{\ifmmode{e^+e^-}\else{$e^+e^-$}\fi}
\def\qq{\ifmmode{q\bar{q}}\else{$q\bar{q}$}\fi}
\def\QQ{\ifmmode{Q\bar{Q}}\else{$Q\bar{Q}$}\fi}
\def\sthw{\ifmmode{\sin(\theta_W}\else{$\sin(\theta_W)$}\fi}
\def\s2thw{\ifmmode{\sin^2(\theta_W)}\else{$\sin^2(\theta_W)$}\fi}
\def\cthw{\ifmmode{\cos(\theta_W)}\else{$\cos(\theta_W)$}\fi}
\def\c2thw{\ifmmode{\cos^2(\theta_W)}\else{$\cos^2(\theta_W)$}\fi}
\newread\epsffilein    
\newif\ifepsffileok    
\newif\ifepsfbbfound   
\newif\ifepsfverbose   
\newif\ifepsfdraft     
\newdimen\epsfxsize    
\newdimen\epsfysize    
\newdimen\epsftsize    
\newdimen\epsfrsize    
\newdimen\epsftmp      
\newdimen\pspoints     
\def\epsfbox#1{\global\def\epsfllx{72}\global\def\epsflly{72}%
   \global\def\epsfurx{540}\global\def\epsfury{720}%
   \def\lbracket{[}\def\testit{#1}\ifx\testit\lbracket
   \let\next=\epsfgetlitbb\else\let\next=\epsfnormal\fi\next{#1}}%
\def\epsfgetlitbb#1#2 #3 #4 #5]#6{\epsfgrab #2 #3 #4 #5 .\\%
   \epsfsetgraph{#6}}%
\def\epsfnormal#1{\epsfgetbb{#1}\epsfsetgraph{#1}}%
\def\epsfgetbb#1{%
\openin\epsffilein=#1
\ifeof\epsffilein\errmessage{I couldn't open #1, will ignore it}\else
   {\epsffileoktrue \chardef\other=12
    \def\do##1{\catcode`##1=\other}\dospecials \catcode`\ =10
    \loop
       \read\epsffilein to \epsffileline
       \ifeof\epsffilein\epsffileokfalse\else
          \expandafter\epsfaux\epsffileline:. \\%
       \fi
   \ifepsffileok\repeat
   \ifepsfbbfound\else
    \ifepsfverbose\message{No bounding box comment in #1; using defaults}\fi\fi
   }\closein\epsffilein\fi}%
\def\epsfclipoff{\def\epsfclipstring{\ifepsfdraft\space clip\fi}}%
\epsfclipoff
\def\epsfsetgraph#1{%
   \epsfrsize=\epsfury\pspoints
   \advance\epsfrsize by-\epsflly\pspoints
   \epsftsize=\epsfurx\pspoints
   \advance\epsftsize by-\epsfllx\pspoints
   \epsfxsize\epsfsize\epsftsize\epsfrsize
   \ifnum\epsfxsize=0 \ifnum\epsfysize=0
      \epsfxsize=\epsftsize \epsfysize=\epsfrsize
      \epsfrsize=0pt
     \else\epsftmp=\epsftsize \divide\epsftmp\epsfrsize
       \epsfxsize=\epsfysize \multiply\epsfxsize\epsftmp
       \multiply\epsftmp\epsfrsize \advance\epsftsize-\epsftmp
       \epsftmp=\epsfysize
       \loop \advance\epsftsize\epsftsize \divide\epsftmp 2
       \ifnum\epsftmp>0
          \ifnum\epsftsize<\epsfrsize\else
             \advance\epsftsize-\epsfrsize \advance\epsfxsize\epsftmp \fi
       \repeat
       \epsfrsize=0pt
     \fi
   \else \ifnum\epsfysize=0
     \epsftmp=\epsfrsize \divide\epsftmp\epsftsize
     \epsfysize=\epsfxsize \multiply\epsfysize\epsftmp
     \multiply\epsftmp\epsftsize \advance\epsfrsize-\epsftmp
     \epsftmp=\epsfxsize
     \loop \advance\epsfrsize\epsfrsize \divide\epsftmp 2
     \ifnum\epsftmp>0
        \ifnum\epsfrsize<\epsftsize\else
           \advance\epsfrsize-\epsftsize \advance\epsfysize\epsftmp \fi
     \repeat
     \epsfrsize=0pt
    \else
     \epsfrsize=\epsfysize
    \fi
   \fi
   \ifepsfverbose\message{#1: width=\the\epsfxsize, height=\the\epsfysize}\fi
   \epsftmp=10\epsfxsize \divide\epsftmp\pspoints
   \vbox to\epsfysize{\vfil\hbox to\epsfxsize{%
      \ifnum\epsfrsize=0\relax
        \includegraphics{\ifepsfdraft}%
      \else
        \epsfrsize=10\epsfysize \divide\epsfrsize\pspoints
        \includegraphics{\ifepsfdraft}%
      \fi
      \hfil}}%
\global\epsfxsize=0pt\global\epsfysize=0pt}%
{\catcode`\%=12 \global\let\epsfpercent=
\long\def\epsfaux#1#2:#3\\{\ifx#1\epsfpercent
   \def\testit{#2}\ifx\testit\epsfbblit
      \epsfgrab #3 . . . \\%
      \epsffileokfalse
      \global\epsfbbfoundtrue
   \fi\else\ifx#1\par\else\epsffileokfalse\fi\fi}%
\def\epsfempty{}%
\def\epsfgrab #1 #2 #3 #4 #5\\{%
\global\def\epsfllx{#1}\ifx\epsfllx\epsfempty
      \epsfgrab #2 #3 #4 #5 .\\\else
   \global\def\epsflly{#2}%
   \global\def\epsfurx{#3}\global\def\epsfury{#4}\fi}%
\def\epsfsize#1#2{\epsfxsize}
\let\epsffile=\epsfbox

\newcommand{\seqn}{\begin{equation}}
\newcommand{\eeqn}{\end{equation}}
\newcommand{\seqna}{\begin{eqnarray}}
\newcommand{\eeqna}{\end{eqnarray}}

\newcommand{\ttbar}{t \bar{t}}

\newcommand{\htop}{H_{Top}}
\newcommand{\hdata}{H_{Data}}

\newcommand{\ztt}{Z \rightarrow \tau \tau}

\newcommand{\tem}{\ttbar \rightarrow e\mu}
\newcommand{\pbi}{pb^{-1}}

\newcommand{\chisq}{\chi^{2}}
\newcommand{\chitop}{\chi^{2}_{Top}}

\newcommand{\chidata}{\chi^{2}_{Data}}
\newcommand{\chiz}{\chi^{2}_{Z}}

\newcommand{\met}{\mbox{$\rlap{\kern0.25em/}E_T$}}
\newcommand{\metc}{\mbox{$\rlap{\kern0.25em/}E_T^{\rm cal}$}}

\begin{document}
\title{Search for the Top Quark at D\O\ using Multivariate Methods}
\author{Pushpalatha C. Bhat \\for the D\O\ Collaboration}
\address{Fermi National Accelerator Laboratory,\\
Batavia, IL 60510, U.S.A.}
\maketitle
\begin{abstract}

We  report on  the search  for the  top  quark  in  $\ppbar$   collisions at
the
Fermilab  Tevatron   ($\sqrt{s}=1.8$  $TeV$) in  the  di-lepton and
lepton+jets
channels using multivariate methods. An  $H$-matrix analysis of the $e \mu$
data
corresponding to an integrated luminosity of $13.5 \pm 1.6$ $pb^{-1}$ yields
one
event  whose   likelihood to  be a  top  quark  event,  assuming
$m_{top}=180$
GeV/c$^2$, is ten times more than that of $WW$ and eighteen times more than
that
of $\ztt$. A neural network analysis of the $e+$jets channel using a data
sample
corresponding to an  integrated  luminosity of $47.9\pm  5.7$ $pb^{-1}$ shows
an
excess of events  in the  signal region and yields a  cross-section for
$\ttbar$
production of $6.7\pm  2.3$ (stat.) $pb$, assuming a  top mass of 200
GeV/c$^2$.
An analysis of the $e+$jets data using the probability density estimation
method
yields a cross-section that is consistent with the above result.

\end{abstract}

\section*{INTRODUCTION}

    The top quark that remained elusive for over a decade and a half has
finally
been  observed by both  the CDF and  D\O\   collaborations\cite{CDF,D0}. The
top
quark events have been observed in the  di-lepton and lepton+jets decay modes
of
$\ttbar$ pairs  produced in $\ppbar$  collisions at  $\sqrt{s}=1.8$ $TeV$ at
the
Fermilab Tevatron. The collaborations have used conventional analysis methods
to
optimize  cuts on  kinematic  variables  together  with the  tagging  of the
$b$
quarks,  to  discriminate top  quark events  from background.   The
conventional
analysis methods do not exploit correlations  amongst the variables on which
the
cuts are  applied and  thus may suffer   a loss in  signal  efficiency. The
D\O\
collaboration  has been  applying multivariate  methods such as  the
$H$-matrix,
probability  density estimation  (PDE) and  neural networks for  identifying
top
quark events   \cite{DPF,Hannu2},  to  improve the  signal  efficiency.  In
this
paper, we describe the multivariate  methods used, we present an analysis of
the
channel  $\ttbar  \rightarrow e\mu$ and  report on  the  measurement of
$\ttbar$
production    cross-section from a  study  of the  channel  $\ttbar
\rightarrow
e+$jets.

\section*{Multivariate Classifiers}

    A  classifier is any  procedure  that  assigns  objects to  classes.  In
the
present context, a classifier would  separate signal events from the
background.
The time-honored  conventional  classification methods of  examining
uni-variate
(1-dimensional)  and  bi-variate (2-dimensional)  distributions  of variables
to
optimize  cuts for  separating signal  and  background events do  not in
general
provide the  maximum possible  discrimination   when  correlations exist
between
variables.  Multivariate  classifiers which fully  exploit the correlations
that
exist among several variables provide a  discriminating boundary  between
signal
and background in multi-dimensional space that can yield discrimination close
to
the theoretical maximum (Bayes' limit \cite{Bayes}).

    In the   multivariate  approach,  one  encodes  each  event as a  point in
a
multi-dimensional  space, called {\it  feature space}, corresponding to a
vector
$x$ of feature  variables such as  electron $E_T$  ($E_T^{e}$),  neutrino
$E_T$,
($\met $), $H_T$  ($\Sigma E_T  (jets)$), {\it etc.} This  feature space is
then
mapped into  a one  or a   few-dimensional output  space in such  a way that
the
signal and  background vectors are  mapped onto different  regions of the
output
space. The aim of the  multivariate methods is to  reduce  the dimensionality
of
the  problem without  losing  information in  the  process.  The  optimal way
to
partition the feature space into signal  and background regions is to choose
the
mapping to be  the Bayes  discriminant  function.  Each cut on  the value of
the
function corresponds to a  discriminating boundary  in feature space. The
Bayes
discriminant  function is  simply the ratio of  the probability  $P(s|x)$ that
a
given  event is  a  signal  event and  the  probability   $P(b|x)$ that  it is
a
background event. It is written as
\begin{equation}
 R(x)  = {P(s|x) \over P(b|x)} = {P(x|s)P(s) \over P(x|b)P(b)}.
\end{equation}
 The quantities
$P(x|s), P(x|b) $ are  the likelihood
functions for the signal and background,
respectively (hereafter denoted as $f(x)$ with or without appropriate
subscript).
The ratio of the prior probabilities $P(s) \over P(b)$ is the ratio of the
signal and background cross-sections. Some multivariate classifiers
approximate the likelihood functions while the
neural network classifier arrives at the Bayesian probability for the signal,
$P(s|x)$,
without calculating the likelihood functions for each class
separately.
The three classifiers being used by D\O\ are described in the following
sections.

\subsection*{$ H-$matrix Method}

    This is the familiar covariance matrix method which is also known as
the Gaussian Classifier.
It was introduced in  the 1930s\cite{Fisher,Maha} as a tool for
discriminating one class of
feature vector $x$
from another.  The vector $x$ is assumed to be distributed
according to a multivariate Gaussian with covariance matrix $M$
and mean $\bar{x}$.  The likelihood function is therefore,
\begin{equation}
 f(x) = A\cdot\exp\{-{1 \over 2} {\sum_{i,j}}(x_i-\bar{x}_i)^T (M^{-1})_{ij}
(x_j-\bar{x}_j)\} \equiv A\cdot\exp(- \chi^{2})
\end{equation}
where $(M^{-1})_{ij}$ is the $H$-matrix.  Fisher\cite{Fisher} showed that
the optimal way to separate two overlapping multivariate Gaussian
distributions with a common covariance matrix but with different means
$\bar{x}_s$ and $\bar{x}_b$ is to cut on the function,
\begin{equation}
F = {1 \over 2}( \chi_b ^2 - \chi_s ^2);
\end{equation}
F is called the Fisher linear discriminant function.  If the two
distributions have different correlation matrices, one can
introduce a more general Gaussian classifier\cite{Kendall}, where the
$\chisq$ values are calculated using the
corresponding
$H$-matrices as
well as mean values for the signal and background classes.  We note that
this method is useful even
when the distribution of $x$ is non-Gaussian.

    The Bayes discriminant function $R(x)$ can be written in terms of the
Fisher variable F as $R(x)=\exp(F)$ when
 $P(s)=P(b)$.

\subsection*{Probability Density Estimation (PDE) Method}

    In the PDE method \cite{Hannu1}, the likelihood functions or the
probability density functions (pdf's) are
approximated by summing over
multivariate kernel functions with one kernel function centered at
each data point for the two classes of events.
The  expression for the likelihood
function is,
\begin{equation}
f(x) = {1 \over N_{events} h_{1}...h_{d}} {\sum_{i=1}^{N_{events}}}
\prod_{j=1}^d K({x_j-x_{ij} \over h_j}),
\end{equation}
where the kernel function K is chosen to be a Gaussian.
    The $j^{th}$ variable is denoted by $x_j$ and, $x_{ij}$ denotes
the $j^{th}$ variable of
the $i^{th}$ event.  By appropriate transformation, the variables $x_j$ are
rendered uncorrelated within the signal and background classes.
 The quantity
$h_j$ is the $j^{th}$ smoothing
    parameter.  We use a single ``global'' smoothing parameter $h$ defined by
\begin{equation}
h_{sj}=h\sigma_{sj}, \hspace{0.5in}  h_{bj}=h\sigma_{bj},
\end{equation}
where $\sigma_{sj}$ and $\sigma_{bj}$ are the estimated
standard deviations of the $j^{th}$
 variable for the signal and background classes,
respectively.  The value of the smoothing parameter $h$ is set by maximizing
the
signal to background ratio (S/B) at the required signal efficiency.

The discriminant function we use in the PDE method is
\begin{equation}
D(x)={f_s(x) \over {f_s(x)+f_b(x)}}
\end{equation}
where $f_s(x)$ and $f_b(x)$ are the pdf's for signal and background classes
of events, respectively.  The function $D(x)$ approximates the Bayesian
probability
for the signal $P(s|x)$.  When $P(s)=P(b)$,
the Bayes
discriminant function becomes $R(x) = D(x)/(1-D(x))$.
\subsection*{Neural Networks}

\begin{figure}[htp]
\centerline{\epsfysize=2.50in \epsffile{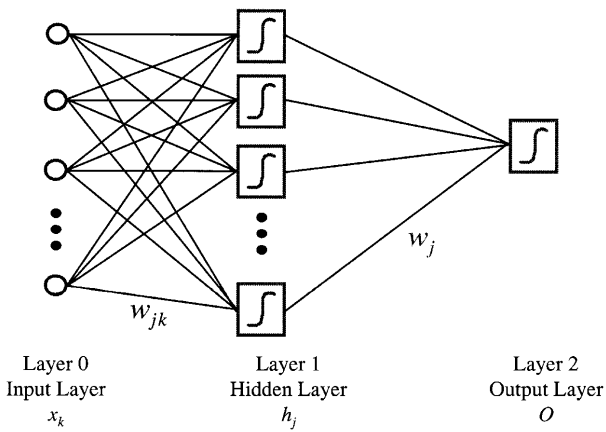}}
\caption{A feed-forward neural network with one hidden layer.}
\label{fig:FFNN}
\end{figure}

    Artificial neural networks provide a powerful new paradigm for event
classification. The most
    commonly used architecture in classification problems is the
multi-layer perceptron or feed-forward neural
    network.  In Fig.
   \ref{fig:FFNN},
 we show the representation of a three layer feed-forward neural network
with one hidden layer.
The nodes in the input layer
    correspond to the components $x_k$ of the feature vector $x$,
and the
    output layer has a single node commonly used in binary classification
problems. The network
    builds an internal representation of the mapping
of the feature space into the output space.  The output of
    the network is given by
\begin{equation}
O(x) = g({{\sum_j} w_{j} g({\sum_k} w_{jk}x_k+\theta_j})+\theta),
\end{equation}
where the ``weights''  $w_{jk}$ and $w_j$ and, the ``thresholds''
$\theta_j$ and $\theta$
 are parameters
that are adjusted during the ``training'' process.
 The quantity g is a non-linear ``transfer''
function of
the form $g(y)=1/(1+e^{-2y})$.  (Use of such transfer functions enables
the mapping of any real function \cite{Blum}.)
The parameters are determined by minimizing the mean square error between
the actual
output $O^p$ and the desired output $t^p$
\begin{equation}
E={1 \over 2N_p} {\sum_{p=1}^{N_p}} {(O^p-t^{p})}^2
\end{equation}
with respect to the parameters.  Here p denotes a feature vector or
{\it pattern}.
Once the parameters are determined using a large number of signal and
background
events the network can be used to classify  events.
    It has been shown\cite{Ruck} that the feed-forward neural network when
trained as a classifier
using  the  back-propagation algorithm for updating the parameters, yields
an output that approximates
the Bayesian probability for the signal
{\it i.e.,} $ O(x)=P(s|x)$.
(This assumes $t^p$ is 1 for signal and 0 for background.)  The Bayes
discriminant in terms of the network output will be
$R(x) = {O(x)/(1-O(x))}$.

\section*{Analysis of the D\O\ Data}

We have applied the multivariate methods to the analysis of the $\tem$ and
$\ttbar \rightarrow e+$jets
channels.  The on-line trigger selection, off-line electron and
muon identification criteria and description of variables used here
can be found elsewhere\cite{Bantly,PRD}.

The $e\mu$ data analysed here correspond to an integrated luminosity of
 13.5$\pm$1.6
$\pbi$. The overall trigger efficiency is about
(90$\pm$7)$\%$  for $m_{top}$= 180 GeV/c$^2$ and
varies slightly  with top mass.
In the off-line selection, before analysis with the $H$-matrix method, we
apply
loose electron and muon  identification  criteria and require
$E_T^e >$11 GeV and P$_T^{\mu}>$11 GeV/c and at least two reconstructed jets
($E_T^{jet} >8$ GeV). Using the $H$-matrix method we have examined
 the signal to
background  ratio with respect to  $WW \rightarrow e\mu$ and
$\ztt \rightarrow e\mu$
(the dominant backgrounds).

We have used the PDE and neural network methods to analyze $e+$jets data
 corresponding to
an integrated luminosity of 47.9$\pm$5.7 $\pbi$.
The dominant background to the $\ttbar \rightarrow e+$jets channel is from the
 QCD production of $W$+multi-jets
where the signature is the same as that of the
signal, {\it viz.} a high $P_T$ electron, high $\met $,
arising from the leptonic decay of the W boson,
and several jets.  In addition, we have background from
QCD multi-jet events where one of the jets is mis-identified as an electron
and the event also has a high $\met $
from mis-measurements as well as neutrinos from
any
heavy flavor decays. We refer to this background as QCD
fakes.
The two backgrounds in our data prior to the multivariate analyses
are estimated directly from data.  The QCD fake background is estimated by the
 joint
probability of multi-jet events having $\met$ larger than the cut applied and
 a jet
being mis-identified as an electron.  The $W$+jets background is estimated
 using Berends' scaling \cite{Berends}.
 The inclusive jet multiplicity data (after subtracting the QCD fakes
 background) is fitted for this
``jet-scaling'' allowing for
contribution from
top quark events.  These background numbers are multiplied by
the fraction of  events surviving the multivariate cuts to get the final
 background estimates.

\subsection*{$H$-matrix Analysis of e$\mu$ Data}
\begin{figure}[htp]
\centerline{\epsfysize=3.0in \epsffile{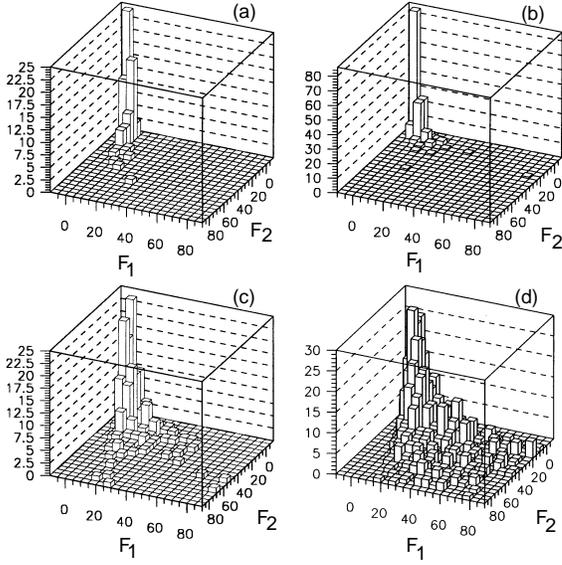}}
\caption{$F_1$ vs $F_2$ from $H$-matrix analysis of $e\mu$ channel
for (a) D\O\ data ($\int Ldt$=13.5 $\pbi$),
 (b) $\ztt$ (($\int Ldt$=3.1 $fb^{-1}$),
(c) $WW$ ($\int Ldt$=22.3 $fb^{-1}$) and (d) $\ttbar 180$
($\int Ldt$=20.1 $fb^{-1}$) samples.}
\label{fig:HM}
\end{figure}

 We have chosen the following feature vector
$x$=($E_T^e$, P$_T^{\mu}$, $E_T^{jet1}$,$E_T^{jet2}$, $\metc $, $H_T$,
 $M_{e\mu}$, $\Delta\phi_{e\mu}$)
to build the
signal and background $H$-matrices, where $\metc $ is the $\met $ in the
 calorimeter, $M_{e\mu}$ is the
invariant mass of the two
leptons and
$\Delta\phi_{e\mu}$ is the azimuthal angle between the two leptons.
We have built the signal $H$-matrix $\htop$ using 180 GeV $\ttbar$
ISAJET Monte Carlo events ($\ttbar$180)
 processed through the D\O\  detector  simulation program. Since in the
$e\mu$ data  we expect very few $\ttbar \rightarrow  e\mu$ events, we have
chosen to use  the data to model the background.
The
data consist largely  of QCD $b\bar{b}$ events and
 $ W \rightarrow \mu$+jets+e$^{\prime}$
events (where e$^{\prime}$ denotes a jet that fakes an electron). We have
 considered  the
$\ztt$ background separately to get better rejection against
that background.
We define two Fisher discriminant functions
\begin{equation}
F_1 = {1\over2}(\chidata - \chitop), \hspace{.50in}
F_2 = {1\over2}(\chiz - \chitop).
\end{equation}
where,
\begin{equation}
\chidata = {\sum_{i,j}}(x_i-\bar{x_i})^T \hdata (x_j-\bar{x_j}), \hspace{0.2in}
\chiz = {\sum_{i,j}}(x_i-\bar{x_i})^T H_{Z} (x_j-\bar{x_j}),
\end{equation}
\begin{equation}
\chitop = {\sum_{i,j}}(x_i-\bar{x_i})^T \htop (x_j-\bar{x_j}),
\end{equation}
where
$\hdata$ and $H_Z$ are the  background  $H$-matrices built using data and
 $\ztt$ Monte Carlo
events, respectively.
The $\chisq $ values, $F_1$ and $F_2$ are determined for signal, backgrounds
 and for data.
In Fig. \ref{fig:HM} we show the lego plots of  $F_1$  {\it vs} $F_2$  for
 each of the samples. By applying
the cuts $F_1 >$ 15 and $F_2 >$3 we have 16\%, 22\% and 25\% efficiency for
 top events with top masses of 140, 160
and 180 GeV/c$^2$, respectively.
The only event that survives the cuts is the same as that found in the
 conventional analysis \cite{PRL1}.
This event lies in a region of phase space where the
the signal to background ratio (S/B)
 is about 18 with respect to $\ztt$ and 10 with respect to $WW$ for a
 180 GeV/c$^2$ top quark.

\subsection*{PDE Analysis of $e+$ jets  Data  }

\begin{figure}[htp]
\centerline{\epsfysize=2.7in \epsffile{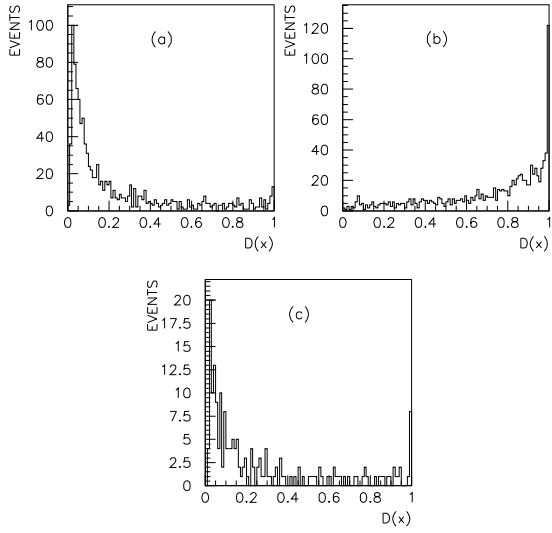}}
\caption{The PDE discriminant function for (a)background,
 (b)$\ttbar$180 and for (c)D\O\ data.}
\label{fig:pdefig}
\end{figure}

The PDE method has been applied to the $e+\geq 3$jets data
\cite{Hannu2}. The selection
criteria  used are $E_T^e >$20 GeV, $\met >$ 20 GeV and at least
3 jets with $E_T>$ 15 GeV. These five transverse  energies define our feature
 vector
in the analysis.   The two backgrounds  are  combined  in the
ratio estimated as in the conventional   analysis and are treated as a
single  background to build the pdf.
 Figure \ref{fig:pdefig} shows the distributions of the discriminant function
  $D(x) $
for background events,  180 GeV/c$^2$ top quark events and for   D\O\ data.
Applying a cut of $D(x) > $ 0.8  yields  21  data  events with an
estimated background of 14.0$\pm$1.6 events in 47.9 $pb^{-1}$.
The product of efficiency and branching ratio for $\ttbar$180 events is 3.1\%
(as compared to 1.8\% for conventional analysis).
 The $\ttbar$  cross-section  is calculated to be   4.7$\pm$3.3 (stat.) $pb$,
 in agreement with the results from the conventional analysis .

\subsection*{Neural Network Analysis of $e+ $jets Data  }

\begin{figure}[htp]
\centerline{\epsfysize=2.0in \epsffile{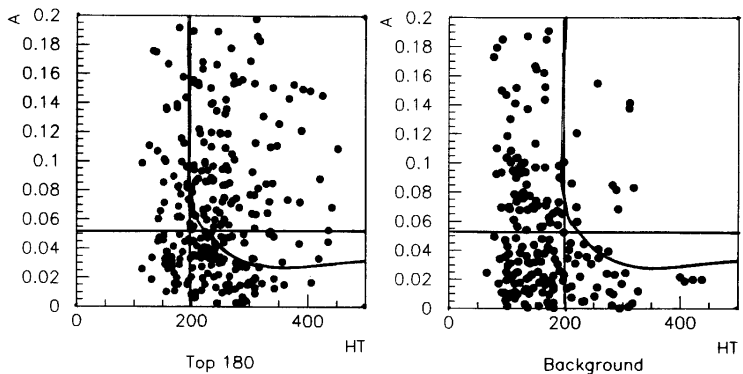}}
\caption{$A$ and $H_T$ scatter plot for signal and background with
 conventional cuts (lines parallel
to the co-ordinate axes) and neural network cut (the contour). }
\label{fig:NN2V}
\end{figure}

 A discussion of a two-variable and a six-variable analysis of the
 $e+\geq 4$jets
data using neural networks has been presented previously\cite{DPF}.
Here we present results of the analyses
including recent data.
The neural network program used here is JETNET 3.0\cite{JETNET}.
We use the $E_T$ of the various measured objects in the event ($E_T^e,
\met , E_T$ of jets), the event shape variable aplanarity
($A$) and the total transverse energy $H_T$ of central jets
(pseudorapidity $|{\eta}|<$2.0) to discriminate the top signal events from the
backgrounds. In our
conventional analysis of $e+\geq 4$jets using non-tagged data we have applied
 selection
cuts of $E_T^e>$20 GeV, $\met>$25 GeV,
$E_T($jet4$)>$15 GeV, $A>$.05 and $H_T>$200 GeV.
(Jets are ordered in decreasing $E_T$; jet4 refers to the jet with fourth
 highest $E_T$.)
 For demonstration
purposes, we compare in  Fig. \ref{fig:NN2V} the conventional cuts on $A$ and
$H_T$ with the contour cut obtained by a simple network with 2 input nodes,
 2 hidden nodes and one
output  node. The $A$ and $H_T$ are used as inputs and the
network is trained on $\ttbar$180 and background events (a mixture of $W$+jets
and QCD fakes combined in the proper ratio).
The contour provides better signal efficiency than the
conventional cuts for the same signal to background ratio.

In order to achieve higher signal efficiency we have relaxed the number of jets
required and have carried out an analysis of the $e+\geq 3$jets data. Also, we
 use a
five-dimensional feature vector $x$=($E_T^e, \met, H_T, A$ , $ E_T($jet3$)$).
For this analysis, we use two different networks to
discriminate against $W + $jets and QCD fake background
separately.  That is, we train one network with $\ttbar$180
as signal and $W+$jets Monte Carlo  events (using the VECBOS event generator)
 as background and
a second network with $\ttbar$180 as signal and QCD fakes
(data events that fail the electron ID cuts)
as background.
We use networks with 5 input nodes (corresponding to the 5-dimensional feature
 vector),
5 hidden nodes in one hidden layer and 1 output node.
We
use 1300 $\ttbar$ events, 1300 $W+$jets events and 590 QCD
fake events for training.  The testing is done on 2400 $\ttbar$ events (which
include the 1300 events used for training), 1300 $W+$jets events and 590 QCD
 fake
events that were used for training.  Training and
testing on the same set of events with the given sample size can
give rise to an uncertainty ($\approx$ 10\%) in the estimated background which
 is included
in the systematic uncertainty.  The target output of the network  $t^p$
during training is set to be  1 for the signal and 0 for the background.


\begin{figure}[htp]
\centerline{\epsfysize=3.1in \epsffile{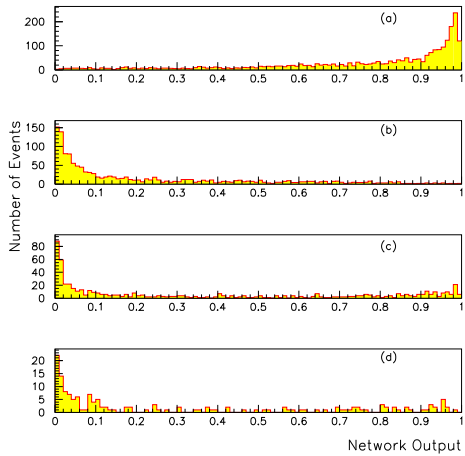}}
\caption{Distributions of the output from the first network for
(a)$\ttbar$180,  (b)$W+$jets (VECBOS), (c)QCD fakes and for  (d)D\O\ data.}
\label{fig:pp_ej1}
\end{figure}

\begin{figure}[htp]
\centerline{\epsfysize=3.0in \epsffile{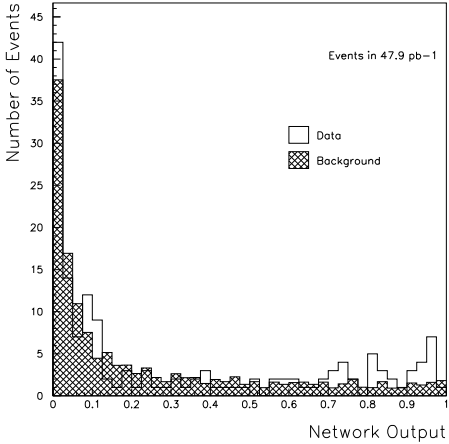}}
\caption{Comparison of outputs from the first network for D\O\ data and
 background }
\label{fig:pp_ej2}
\end{figure}

    Figure \ref{fig:pp_ej1} shows the output of the first network (NN1)
for $\ttbar$180, $W+$jets, QCD fakes and for  D\O\ data.  The
distributions peak close to 1 for signal events and close to
0 for background events, as expected.
In Fig. \ref{fig:pp_ej2}, we show the
output distributions from the first network for data and for the background
($W+$jets and QCD fakes combined) normalized to the number of events
expected in 47.9 $\pbi$.
We have estimated our background to be (80$\pm$11)\% $W+$jets
and (20 $\pm$5)\% QCD fakes. (Errors are statistical only.)
The distributions in Fig. \ref{fig:pp_ej2} are statistically consistent with
 each other in the
background region (NN1 close to 0) and we observe an excess of data events
in the signal region.

\begin{figure}[htp]
\centerline{\epsfysize=3.0in \epsffile{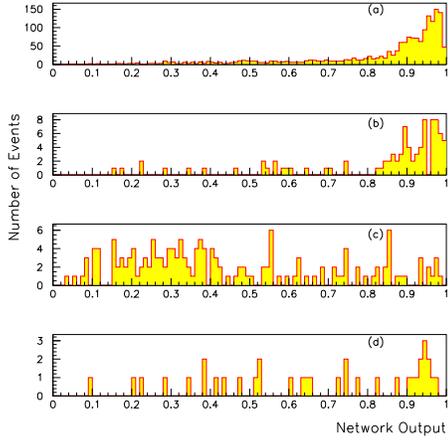}}
\caption{Output of the second network (NN2) after requiring NN1$>$0.7 for
(a)$\ttbar$180,  (b)$W+$jets (VECBOS), (c)QCD fakes and for  (d)D\O\ data.}
\label{fig:pp_ej3}
\end{figure}

For the most part, the kinematic distributions of QCD fakes are similar to
$W+$jets and, therefore, all but a small part of QCD fakes can be
rejected with the first network.  However, to get better rejection of QCD
fakes, we process all samples through the second network.  In Fig.
 \ref {fig:pp_ej3} we show the output of the second network (NN2)
for signal, background and data events which satisfy the cut NN1
$>$0.7.  Applying a cut NN2$>$0.5, we get a factor of three more
reduction in the QCD fake background.

\begin{figure}[htp]
\centerline{\epsfysize=3.5in \epsffile{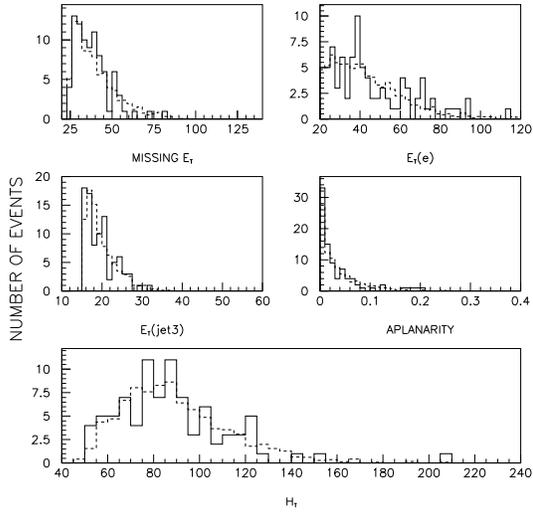}}
\caption{Distributions of input variables compared for data(solid histograms)
 and combined backgrounds
(dashed histograms)
after applying cuts  NN1$<$0.4
and NN2$<$0.4 (anti-Top cuts).}
\label{fig:5vdis}
\end{figure}

We examine the distributions of the five input variables for data and
 background in the region
NN1$<$0.4, NN2$<$0.4
noting that only about 5\% of the $\ttbar$180
events lie in that region.
Given that   the events in the region are mostly background
we can check if our background modeling is correct.
The distributions for data and the combined background are
compared in Fig. \ref{fig:5vdis}.  There is good agreement between
data  and the background model.

Applying the cuts
NN1$>$0.7 and
NN2$>$0.5  yields 25 candidate events with an estimated  background
of 10.1$\pm$1.5.
This gives an excess over background of 14.9$\pm$5.2 events.
The 25
candidate events found here include most of the
non-tagged and $\mu$-tagged candidate events found by the conventional
 analysis.
The product of efficiency and branching ratios are 4.0\% and 4.6\%
 (compared to
1.8 \% and 2.4\% for conventional analysis)  for $\ttbar$180 and
$\ttbar$200, respectively.
For a top
quark mass of 200 GeV/c$^2$, we obtain a $\ttbar$ production cross-section
of 6.7$\pm$2.3 $pb$. (For $\ttbar$180, the cross-section obtained is
7.8$\pm$2.7 $pb$.)
  The errors quoted are statistical only.  A preliminary estimate of the
systematic uncertainty which includes errors in background estimation and
signal efficiency (prior to multivariate analyses)
and neural network specific uncertainties is
about 30\%.  This is dominated by the first two components and work is in
 progress
to reduce these uncertainties.

\section*{SUMMARY}

    We have applied multivariate  analysis methods to search  for top
quark events in the D\O\ data
and we find a significant excess of events over background.
An H-matrix  analysis of  $e\mu$  data ($\int Ldt$
= 13.5$\pm$1.7 $\pbi$) yields one  candidate event (same as found in the
 conventional analysis)
 which lies in a phase space region where S/B = 10 with respect to
$WW$ and S/B = 18 with respect to $\ztt$. Preliminary results from
a PDE analysis  of the $e+\geq 3$jets data
are consistent  with results from  the conventional  analysis.
A preliminary
neural network analysis  of $e+\geq 3$jets data yields
 $\sigma_{\ttbar}(m_{top}=200 GeV/c^2$)
 = 6.7$\pm$2.3 (stat.) $pb$
in agreement with  our published\cite{D0}
 results( $\sigma_{\ttbar}(m_{top}=200 GeV/c^2$)
= 6.3$\pm$2.2 $pb$).

\section*{ACKNOWLEDGEMENTS}

We thank the Fermilab Accelerator, Computing  and Research Divisions and
 support staffs at the
collaborating institutions for their
contributions to this work.

    This work is supported in part by the U.S. Department of Energy.

\end{document}